# Some features in 4-level generation in LIPLs.

L. Nagli[1], K. Kulikov[1], D. Cheskis[1]
[1]Physics Department, Ariel University, Ariel 40700, Israel

This paper shows that in Laser-Induced Plasma Lasers (LIPL) the collisionally assisted transitions that lead to the inversion population on an upper-generation level $E_{up}$ may be partly forbidden. The spin-orbit coupling may increase the oscillator strength of such transitions. It also demonstrates that collisions between electrons and excited atoms can strongly increase the atoms' energy, creating a population inversion at the $E_{up}$, level, which may lie about 1 eV above the pumped level $E_{pump}$. Examples of oscillator strengths and collisional transition rate estimates are provided using linear-response time-dependent density functional theory (TDDFT) in the Casida formalism.

## 1. Introduction

This paper continues our recent work on four-level generation in Laser-Induced Plasma Lasers (LIPL) [1] [1]. LIPLs can be seen as metal-vapor lasers that operate in normal air conditions [2–11] [2,3,4,5,6,7,8,9,10,11]. We suggested that the population inversion at the upper laser level ($E_{up}$) in a four-level LIPL system can be created by direct, one-step collisions between electrons and excited atoms. This means that the $E_{up}$ level is populated without going through many intermediate energy levels. The proposed mechanism is especially relevant for elements of the 14[th] group, where the energy difference between the pumped level ($E_{pump}$) and the $E_{up}$ level $\Delta E = E_{pump} - E_{up}$ **are positive** and large ( about 1 eV), and there exist many intermediate levels between $E_{pump}$ and the $E_{up}$ levels.

Only allowed collisional transitions were considered when estimating the transition rates coefficients $\beta^{mix}$ in [1], following the theory of Ya. Zel'dovich and Yu. Raizer [12] [12].

In addition, our previous studies have only found cases where the $E_{pump}$ is much higher in energy than the $E_{up}$, which corresponds to the inverse $\beta$-process described in [12]. In this paper, we show, for the first time, that **partly allowed** collisionally-assistant transitions may participate in the creation of the inverse population on the $E_{up}$ level. This proposition is supported by theoretical calculations of the transition oscillator strength $f_{ji}$ and transition rates $A_{ji}$ based on time-dependent density functional theory (TDDFT) [13-16] [13,14,15,16].

We also demonstrate that LIPL's generation can occur with a large **negative** energy gap $\Delta E \sim -1$ eV, meaning that about 1 eV of energy may be **transferred from electrons to atoms in** collisional processes.** This process corresponds to the direct α-type collisions, where excited atoms gain significant energy from electron impacts [12].

## 2. Experimental setup

The experimental setup, plasma conditions, and timing details are described in [1]. Briefly, a transverse pumping scheme was used. The laser-induced plasma (LIP) was created by a Nd:YAG laser at 1.06 μm. After a delay (which depended on the sample under investigation), the plasma was pumped by an Optical Parametric Oscillator (OPO). The OPO beam was tuned to match a resonant transition of the element in the plasma.

The spectral resolution of about 0.02 nm and a time resolution of 1.5 ns were in the UV-VIS spectral range. For IR measurements, the spectral resolution was about 6 nm, and the time resolution was about 1 ms.

## 3. Results and discussion

Fig. 1 shows the actual generation spectra of *Ti* LIPL under pumping at 227.26 nm, corresponding to the $4s^2\ ^3F_4 \rightarrow 4s5p\ ^3G_5$ transition. The corresponding generation scheme with collisionally assistant and radiative transitions is shown in Fig. 2

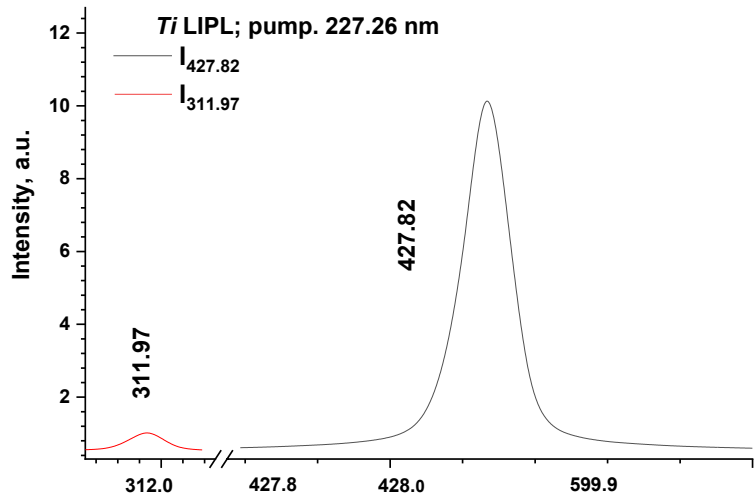

**Fig. 1** *Ti* LIPL actual generation spectra under pumping at 227.26 nm

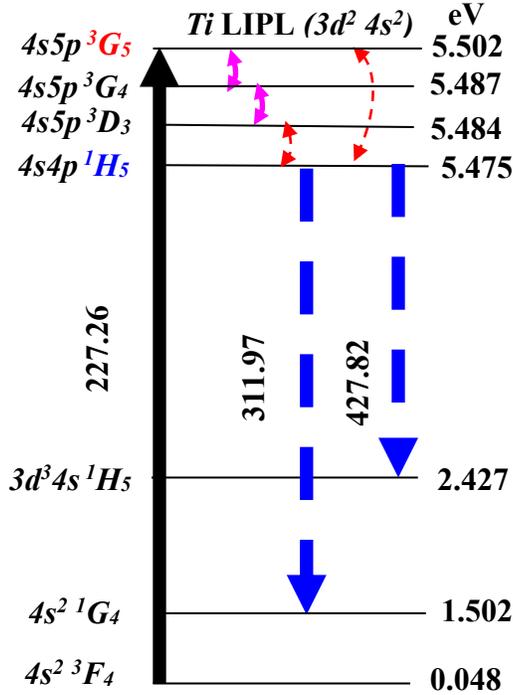

**Fig 2**. Transition scheme of *Ti* LIPLs under pumping at 227.26 nm.
Upward-pointing arrows represent pumping transitions. Violet curved arrows indicate allowed collisional transitions; red curved arrows are partly forbidden transitions. Dashed downward arrows represent lasing transitions.

Lines 311.97 nm, and 472.82 nm correspond to four-level generation from the same 4s4p $^1H_5$ $E_{up}$ level with $\Delta E = 0.027$ eV. According to the scheme in Fig. 2, the population inversion at the $E_{up}$ level for the 311.97 nm and 472.82 nm lines may be achieved via one-step collisional transitions $4s5p\ ^3G_5 \rightarrow 4s4p\ ^1H_5$, or a cascade $4s5p\ ^3G_5 \rightarrow 4s5p\ ^3G_4 \rightarrow 4s5p\ ^3D_3 \rightarrow 4s4p\ ^1H_5$. **It must be noted that jump-like or the last step in the cascade transitions are partly spin-forbidden ($\Delta S \neq 0$).**

Based on this example, the method for estimation of the collisional rates coefficient $\beta^{mix}$ (in cm$^3$/sec), and overall collisional transition rates for cascade transitions $A_C$ and jump-like transitions $A_{jump}$ (in sec$^{-1}$) were developed.

For the numerical evaluation of the parameter $\beta^{mix}$, we follow a comprehensive two-stage approach that combines ground-state and excited-state electronic structure methods. First, the ground-state electronic structure is determined via a fully self-consistent solution of the Kohn-Sham [17][17] equations within Density Functional Theory (DFT). The process begins with an initial guess for the electron density, typically a superposition of atomic densities. At each self-consistent field (SCF) iteration, the effective potential is constructed, incorporating three main contributions: the electron nucleus interaction (represented via pseudopotentials or the projector augmented wave (PAW) method), the classical electrostatic Hartree potential, and the exchange-correlation potential from the chosen DFT functional (here, the hybrid B3LYP functional is used to improve spectral accuracy [18-20]) [18,19,20]

The Kohn-Sham equations take the form:

$$\left(-\frac{h^2}{2m_e}\nabla^2 + V_{eff}[\rho](r)\right)\psi_i(r) = \varepsilon_i\psi_i(r), \quad (1)$$

where $\psi_i(r)$ are the single-particle Kohn-Sham orbitals and $\varepsilon_i$ are the corresponding eigenvalues. The effective potential $V_{eff}[\rho](r)$ includes contributions from three sources:

$$V_{eff}[\rho](r) = V_{ion}(r) + \int \frac{\rho(r')}{|r-r'|}d^3r' + V_{xc}^{B3LYP}[\rho](r), \quad (2)$$

where $V_{ion}$ represents the external ionic potential (typically implemented via pseudopotentials), the second term corresponds to the classical electrostatic (Hartree) potential, and $V_{xc}^{B3LYP}$ is the exchange_correlation potential derived from the *B3LYP* hybrid functional. All calculations were performed with this spin-orbit contribution taken into account, ensuring correct level splitting and modification of the oscillator strength. The equations are solved iteratively using a preconditioned conjugate-gradient algorithm to full convergence, ensuring self-consistency. To capture weakly allowed, spin-forbidden, or strongly correlated transitions, we extend our calculations using linear-response TDDFT in the Casida formalism [13–16]. The resulting $\psi_i(r)$ and $\varepsilon_i$ are then used to compute electric-dipole matrix elements between occupied and unoccupied orbitals, yielding oscillator strengths which, together with physical constants, allow calculation of $\beta_{mix}$ [1]. For each cascade transition (n+1→ n), the rate $A_c$ is

$$A_c = N_e \prod_n^m \beta_{n+1,n}^{mix}, \quad (3)$$

where m is the total number of transitions, n is the transition number and for the direct "jump" transition from level 1 to level 4 is $A_{jump}$

$$A_{jump} = N_e \beta_{1,4}^{mix} \quad (4)$$

Table I summarizes the ordinary numbers *n* of collisional transitions, corresponding energy levels in eV, and its configurations and terms [17].[21] Lust two right colons present the estimated transition average oscillator strength $f_{ji}$ and collisional transitions rate $N_e\beta^{mix}$. For cascade transition, $A_c$ is shown in a separate row.

**Table I.** Parameters for the estimation $f_{ji}$, $A_c$, and $A_{jum}$.

| Ti LIPL (3d²4s) Puming $4s^2\ ^3F_4 \rightarrow 4s5p\ ^3G_5/227.26$ nm; Generation 311.79 nm ($4s4p\ ^1H_5 \rightarrow 4s^2\ ^1G_4$) and 427.82 nm ($4s4p\ ^1H_5 \rightarrow 3d^34s\ ^1H_5$) | | | | |
|---|---|---|---|---|
| n, n' | Energy, (eV) | Configuration/Terms | $f_{ji}$ | $N_e \beta^{mix}$ [sec$^{-1}$] |
| 1 → 2 | 5.5019/($E_{pump}$)/5.48718 | $4s5p\ ^3G_5 \rightarrow 4s5p\ ^3G_4$ | 6.881e-2 | 1.216e11 |
| 2 → 3 | 5.48718/5.484 | $4s5p\ ^3G_4 \rightarrow 4s5p\ ^3D_3$ | 6.885e-2 | 1.217e11 |
| 3 → 4 | 5.484/5.4755 ($E_{up}$) | $4s5p\ ^3D_3 \rightarrow 4s4p\ ^1H_5$ | 2.334e-3 | 4.126e9 |
| 4→ 3→ 2→ 1; $A_c$=0.61 sec$^{-1}$ | | | | |
| 1→ 4 | 5.5019/($E_{pump}$)/5.4755 ($E_{up}$) | $4s5p\ ^3G_5 \rightarrow 4s4p\ ^1H_5$ | 1.29e-3 | $A_{jump}$=2.28e9 |

It is seen that $A_{jump}$ = 2.28e9 sec$^{-1}$ is much higher than $A_c$=0.61 sec$^{-1}$ despite of the presence the partly forbidden collisional transitions ( 4→3 or 1→4 ) and small numbers of the intermediate levels. Forbidden collision transitions are found in other LIPLs, also. For example *V* LIPLs under pumping at 217.7 nm ($4s^2\ ^4F_{7/2} \rightarrow 4s4p\ ^4G_{9/2}$) and generation at 452,96 due to transitions $3d^44p\ ^2G_{7/2} \rightarrow 4s^2\ ^2H_{5/2}$. The transition scheme is presented in Fig 3. It is seen that the $E_{up}$ ($^2G_{7/2}$) level is populated from $^4G_{9/2}$ by partly spin-forbidden transitions.

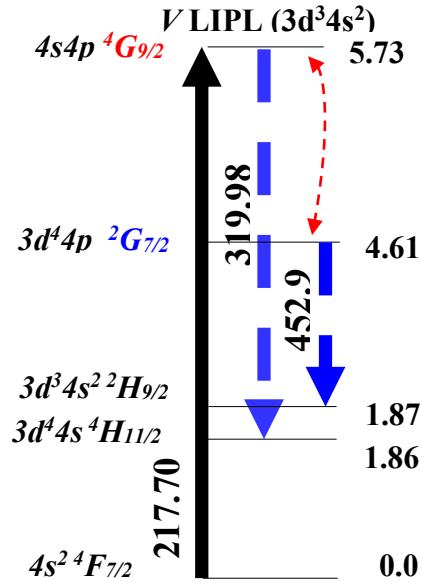

**Fig 3** The actual transition scheme of *V* LIPLs under pumping at 217.70 nm.

Fig. 4 shows the Fe-LIPL generation scheme under pumping at 216.67 nm in which efficient inversion population on $E_{up}$ ($3d^74d\ ^5H_7$) level for 452.96nm generation and on $E_{up}$ ($3d^74d\ ^5H_7$) involves about 1eV **energy transfer to the atoms** for the creation of the inversion population on $E_{up}$ ($3d^74d\ ^5H_7$) level for 452.96 nm generation and $E_{up}$ ($3d^74d\ ^5H_7$) level for 1648.7 nm generation.

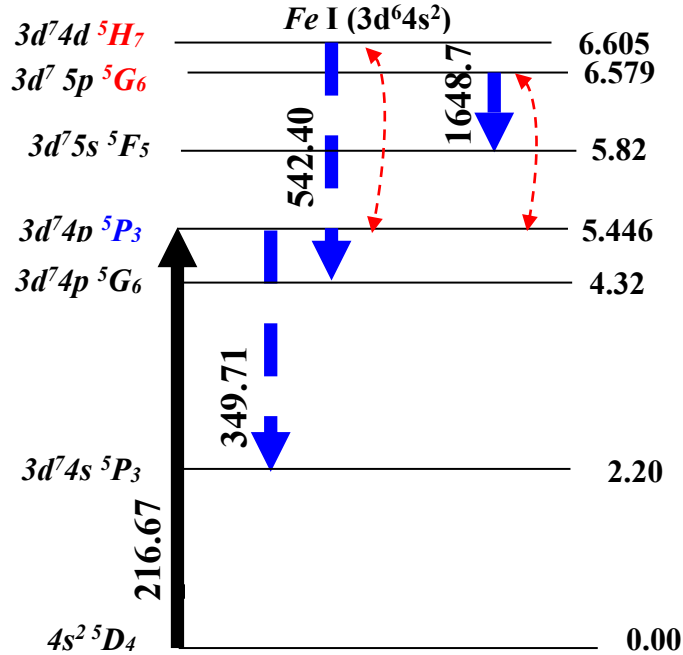

**Fig. 4** The actual part of the *Fe* LIPL generation scheme under pumping at 216.677 nm

In both these cases, the collisional transitions from the $E_{pump}$ level ($3d^74p\ ^5P_3$) to the $E_{up}$ levels ($3d^74d\ ^5H_7$ or $3d^7\ 5p\ ^5G_6$) are partly forbidden.

### 4. Conclusions

In the present paper, it is experimentally demonstrated that, in some cases, the collisional transition that creates population inversion on the $E_{up}$ level for the 4-level generation is partly forbidden due to $\Delta S \neq 0$ or $\Delta J = 2$. The theoretical estimations developed in this manuscript show that despite the partly forbidden transitions involved, its probabilities still are large enough.

It is also shown that in LIPLs it is possible for the large energy transfer from elections to excited atoms in the creation of the inversion population at the $E_{up}$ level.